\documentclass[12pt,a4paper]{article}

\usepackage{rotating}
\usepackage{graphicx}

\title{On the Mathematical Theory of Human Migration: Model of a Migration Channel with a Secondary and a Tertiary Arm}

\author{Roumen Borisov, Nikolay K. Vitanov}
\date{ Institute of Mechanics, Bulgarian Academy of Sciences, Akad. G. Bonchev Str., Bl. 4, 1113 Sofia, Bulgaria}

\begin{document}

\maketitle

\begin{abstract}
We study the motion of substance in a finite channel that belongs to a network. The channel splits to two arms in a node of the network. There is an additional split of the secondary arm. We obtain analytical relationships for the distribution of the substance in the nodes of the channel for the case of stationary regime of the motion of the substance in the arms of the channel. The obtained results are discussed from the point of view of application of the model to migration dynamics: model of motion of migrants in a channel consisting of chains of countries with different probabilities for obtaining permission to stay for the migrants in the different countries of the channel
\end{abstract}

\section{INTRODUCTION}
Complex systems and their nonlinear dynamics are subject of many studies in the last decades \cite{a1} - \cite{vx6} and many results have been obtained
for the behavior and features of the dynamics of social and population complex systems \cite{albert} - \cite{vit6}. 
In the last decades models of flows in networks are much used in the study of different kinds of 
problems, e.g, transportation problems \cite{ff}-\cite{ch1}. In the course of the  years the research 
interest expanded to the : just in time scheduling,  shortest path finding,    facility layout and location,  optimal 
electronic route guidance in urban traffic networks, etc.
Below we shall discuss a model for the motion of a substance
through a network channel in presence of possibility for "leakage" of substance. This model can be applied  to the problem of flow of  a 
substance through a channel with use of part 
of the substance in some industrial process in the nodes of the channel. In addition the model can be applied to the case of  
human  migration flow. 
Human migration  is an actual research topic that is  very important for
taking decisions about economic development of regions of a country \cite{everet} - \cite{borj}.  Human migration is closely connected, e.g., to:  
migration networks; (ii) ideological struggles ; (iii)  waves and statistical distributions in population systems. We note that the probability and deterministic
models of human migration are interesting also from the point of view of applied mathematics \cite{will99} - \cite{rmr07}. 
\par
\begin{figure}[h] \label{fig:net}
	\centering
	\hskip3cm \includegraphics[width=175pt]{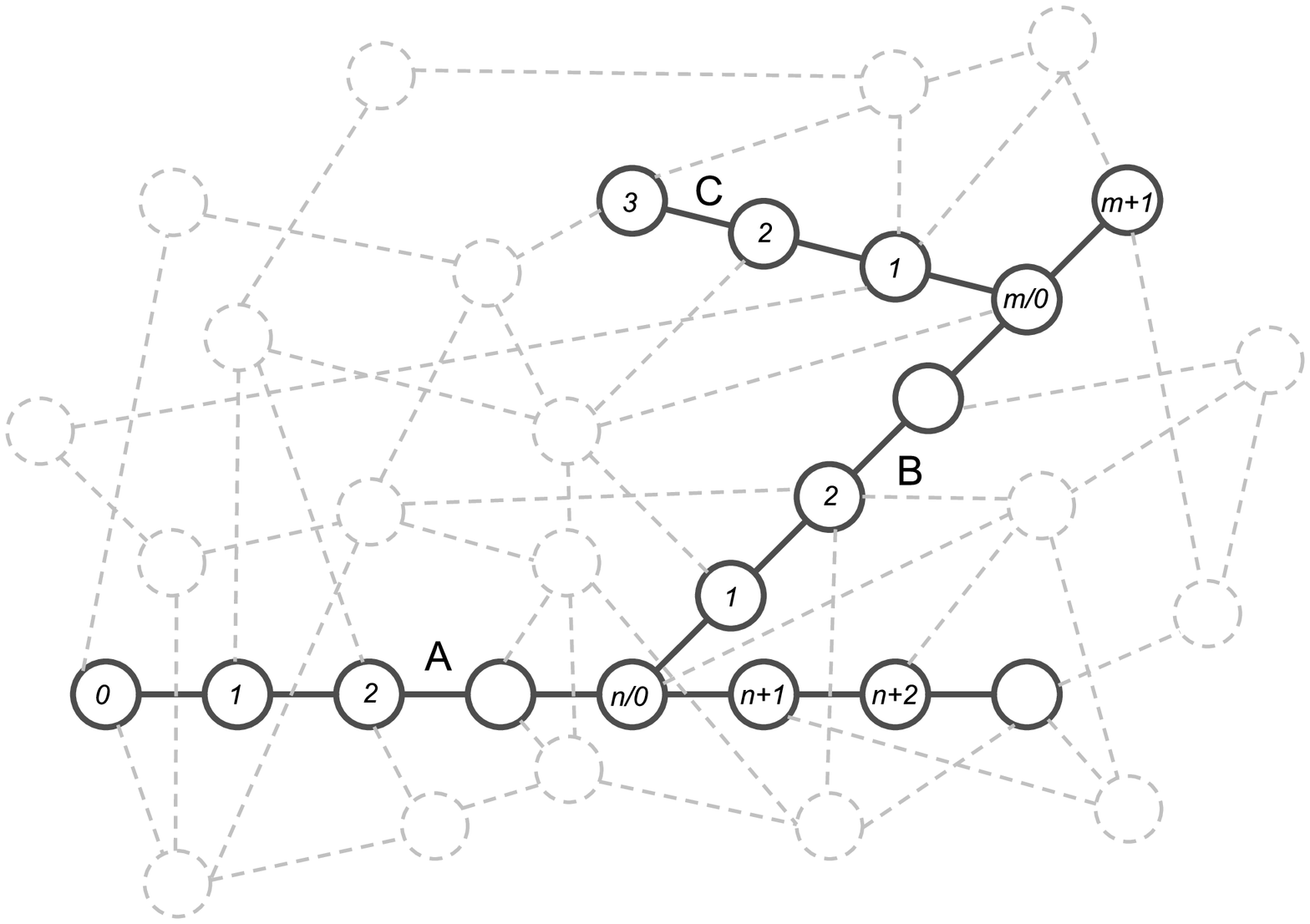} \hskip-2cm
	\includegraphics[width=175pt]{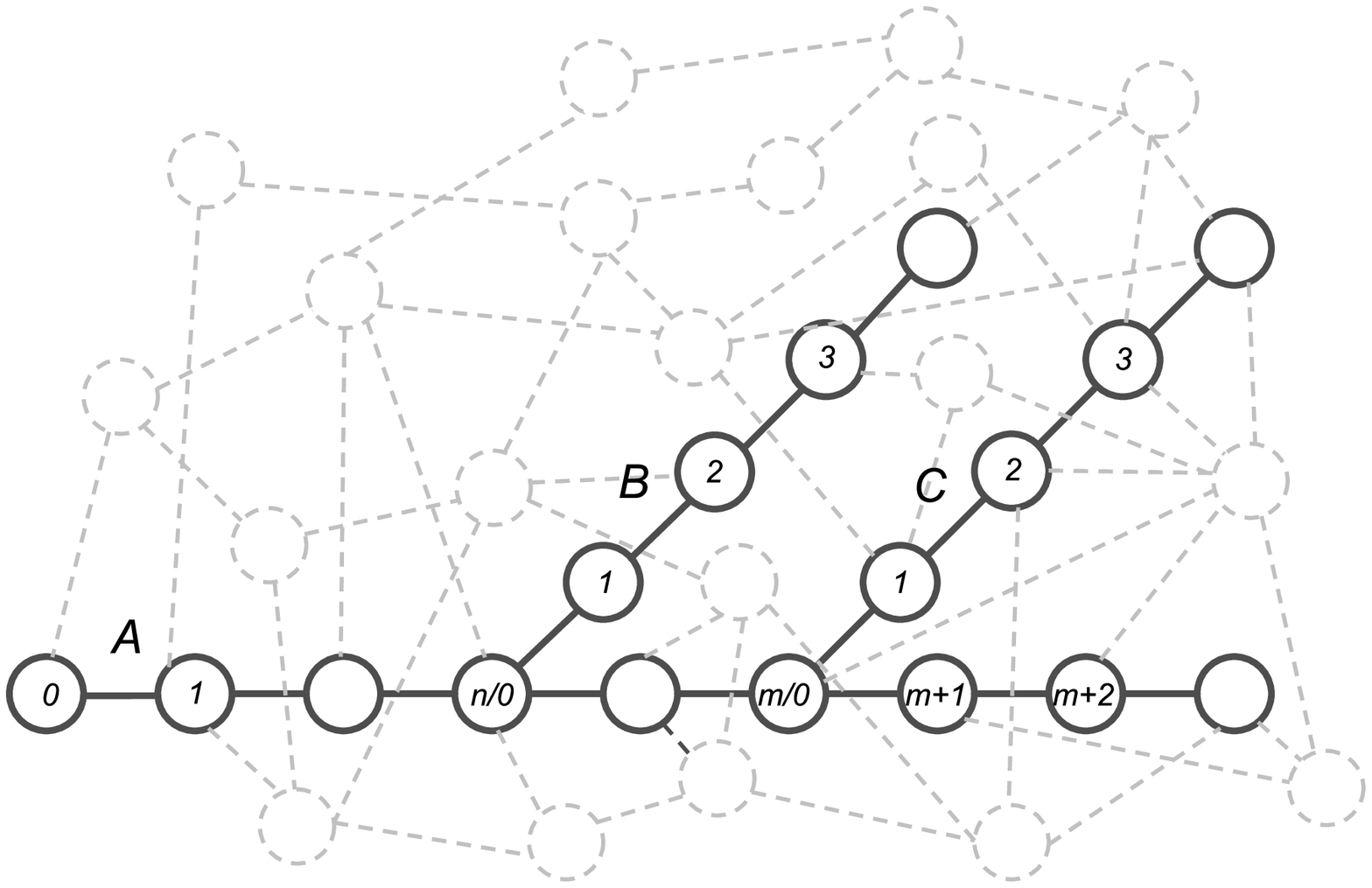}
	\caption{Part of a network and a channel consisting of three arms - $A, B$ and $C$. The nodes of the channel are represented by circles and the edges that connect the nodes are represented by solid lines. At some nodes the channels $A$ and $B$ split to two arms. These nodes are labeled $n/0$ and $m/0$ respectively. $n$ is the number of the splitting node from the point of view of the channel $A$. $m$ is the number of the splitting node for splitting of the arm $C$  from the point of view of the channel $B$ (left hand side network) or from the point of view of the channel $A$ (right hand side network). In this article we shall discuss the left hand side network. The  nodes of the network that are not a part of the studied channel are connected by edges that are represented by dashed lines}
\end{figure}

A specific feature of the study below is that the channel consists of three arms. Depending on the mutual arrangement of these arms, we can consider two different cases. In the first case  (left hand side of Fig. \ref{fig:net}) the channel has a single arm up to the node $n$ where the channel splits into two arms. The second arm of them can be considered as a new single channel up to the node $m$ where it also splits into two arms.
We shall study this channel configuration below. In the second case  (right hand side of Fig.\ref{fig:net}) the channel  has a main arm (arm $A$). The main arm has a single part up to the node $n$ where the channel splits into two arms. The  main arm extension after node $n$ continues and at the node $m$  it also splits into two arms. In this way the channel consists a main arm and two other arms connected to the main one in two different nodes ($n/0$ and $(m/0,\ m>n$). 
\par
There are three special nodes in the above two cases. The first node of the network (called also the entry node and labeled by the number $0$) is the only node of the network where the substance may enter the channel. The second node of the network is the node where the channel splits in two arms  for the first time (node $n/0$). The third special node of the network is the node where  the channel splits in two arms for the second time (node $m/0$). We assume that the substance moves only in one direction along the channel (from nodes labeled by smaller numbers to nodes labeled by larger numbers). The substance may leave the channel at any of the nodes. We shall call this process "leakage" and the "leakage" can be  only in the direction from the channel to the network (and not in the opposite direction). 

\subsection {Stationary regime of functioning of the first and second arms of the discussed channel} \label{sec:num1}
 Each node of the studied channel is connected to the two neighboring nodes of the channel exclusive for the first and $n$-th (and $m$-th respectively) node. The first node  of the  channel is connected only to the neighboring node. We study a model of the motion of substance through such a channel which is an extension of the model discussed in \cite{sg1} and \cite{v1}. We consider each node as a cell (box), i.e., we consider an array of infinite number of cells indexed in succession by non-negative integers. Further 
we assume that an amount $x^{q}$ of some substance  is distributed among the cells in channel $q$ ($q\in\{A,B\}$) and this substance can move from one cell to another cell. The upper index denote the arm of the channel.  Let $x_i^{q}$ be the amount of the substance in the $i$-th cell of the channel $q$. Then
\begin{equation}\label{21}
x^q = \sum \limits_{i=0}^\infty x_ i^q.
\end{equation}
The fractions $y_i^q= x_i^q/x^q$ can be considered as probability values of distribution of a discrete random variable $\zeta$: 
$y_i^q = p(\zeta = i), \ i=0,1, \dots$.
The content $x_i^q$ of any cell may change due to the following 3 processes:
\begin{enumerate}
	\item Some amount $s^q$ of the substance $x^q$  enters the system of cells from the external environment through the $0$-th cell;
	\item Rate $f_i^q$ from $x_i^q$ is transferred from the $i$-th cell into the $i+1$-th cell;
	\item Rate $g_i^q$ from  $x_i^q$  leaks out the $i$-th cell into the external environment.
\end{enumerate}
We assume that the process of the motion of the substance is continuous in the time. Then the process can be modeled mathematically by the system of ordinary differential equations:
\begin{eqnarray} \label{23}
\frac{dx_0^q}{dt} &=& s^q-f_0^q-g_0^q; \nonumber \\
\frac{dx_i^q}{dt} &=& f_{i-1}^q -f_i^q - g_i^q, \ \ i=1,2,\dots.
\end{eqnarray}
There are  two regimes of functioning of the channel: stationary regime and non-stationary regime. What we shall discuss below is the stationary regime of functioning of the channel. In the stationary regime of the functioning of the channel $dx_i^q/dt=0$, $i=0,1,\dots$.  Let us mark the quantities for the stationary case with $^*$. Then from Eqs. (\ref{23}) one obtains 
\begin{equation}\label{24}
f_0^{*q}=s^{*q}-g_0^{*q}; \ \  f_i^{*q}=f_{i-1}^{*q}-g_i^{*q}.
\end{equation}  
This result can be written also as
\begin{equation}\label{25}
f_i^{*q} = s^{*q}- \sum \limits_{j=0}^i g_j^{*q}
\end{equation}
Hence for the stationary case the situation in the channel is determined by the quantities $s^{*q}$ and $g_j^{*q}$, $j=0,1,\dots$.
\par
In this paper we shall assume the following forms of the amount of the moving substances  in Eqs.(\ref{23}) ( $\alpha^q, \beta^q, \gamma_i^q, \sigma$ are constants)
\begin{eqnarray}\label{26}
s^A &=& \sigma x_0^A = \sigma_0^A; \quad s^B = \delta_n^A x_n^A; \quad \sigma_0^A>0, \ \ \delta_n^A > 0  \nonumber \\
f_i^q &=& (\alpha_i^q + \beta_i^q i) x_i^q; \ \ \ \alpha_i^q >0, \ \beta_i^q \ge 0  \nonumber \\
g_i^q &=& \gamma^{*q}_i x_i^q; \ \ \ \gamma_i^{*q} \ge 0 \to \textrm{non-uniform leakage in the nodes},
\end{eqnarray}
where  $\gamma_i^{*q}= \gamma_i^q+ \delta_i^q $. $\gamma_i^{*A}= \gamma_i^A+ \delta_i^A $ describes the situation with the leakages in the nodes of the channel A. We shall assume that $\delta_i^A=0$ for all $i$ except for $i=n$. This means that in the $n$-th node (where the second arm of the channel splits from the first arm of the channel) in addition to the usual leakage $\gamma_i^A$ there is additional leakage of substance given by the term $\delta_n^A x_n^A=s^B$ and this additional leakage supplies the substance that then begins its motion along the channel~$B$. Furthermore we shall assume that  $\delta_i^B=0$ for all $i$ except for $i=m$. This means that in the $m$-th node in channel $B$ in addition to the usual leakage $\gamma_i^B$ there is additional leakage of substance given by the term $\delta_n^B x_n^B$ and this additional leakage supplies the substance that then begins its motion along the channel $C$.
On the basis of all above the model system of differential equations for these two channels becomes
\begin{eqnarray} \label{27}
\frac{dx_{0}^q}{dt}&=&s^q-\alpha_0^q x_0^q-\gamma_0^{*q}x_0^q; \ \ s^A=\sigma x_0^A = \sigma_0^A; \quad s^B = \delta_n^A x_n^A  \nonumber \\
\frac{dx_{i}^q}{dt}&=&[\alpha_{i-1}^q+(i-1)\beta_{i-1}^q]x_{i-1}^q-(\alpha_{i}^q+ i \beta_{i}^q+\gamma_{i}^{*q})x_{i}^q;\ \ i=1,2,\dots 
\end{eqnarray}
\par
Below  we shall discuss the situation in which the stationary state is established in the entire channel (in the first and second channels respectively). Then $dx_0^q/dt=0$ from the first of the Eqs.(\ref{27}). Hence 
\begin{equation}\label{eq28}
x_0^q=\frac{s^q}{\alpha_0^q+\gamma_0^q} \ \ .
\end{equation}
For the channel $A$ it follows that $\sigma_0^A = \alpha_0^A + \gamma_0^A$. This means that $x_0^A$ (the amount of the substance in the $0$-th cell of the channel $A$) is free parameter. For the channel $B$,  $dx_0^k/dt=0$ follows that 
\begin{equation}\label{29}
x_0^{*B} = \frac{\delta_n^A x_n^{*A}}{\alpha_0^B + \gamma_0^B}\ \ .
\end{equation}
In principle the solution of Eqs.(\ref{27}), $i=1,2,\dots$ is
\begin{eqnarray}\label{210}
x_i^A &=& x_i^{*A} + \sum \limits_{j=0}^i b_{ij}^A \exp[-(\alpha_j^A + j \beta_j^A + \gamma_j^{*A})t];\nonumber \\
x_i^B &=& x^{*B}_i+\sum_{j=0}^{i} b^{B}_{ij}\exp[-(\alpha_j^B + j \beta_j^B + \gamma_j^{*B})t]+ \nonumber \\
&&
\sum_{j=1}^{n} c_{ij}\exp[-(\alpha_j^A + j \beta_j^A + \gamma_j^{*A})t]  
\end{eqnarray}
where $x_i^{*q}$ is the stationary part of the solution. For $x_i^{*q}$ one obtains the relationship (just set $dx^q/dt = 0$ in the second of Eqs.(\ref{27}))
\begin{equation}\label{211}
x_i^{*q} = \frac{\alpha_{i-1}^q + (i-1) \beta_{i-1}^q}{\alpha_i^q + i \beta_i^q + \gamma_i^{*q}} x_{i-1}^{*q}, \ i=1,2,\dots
\end{equation}
The corresponding relationships for the coefficients $b_{ij}^q$ are ($i=1,\dots$):
\begin{equation}\label{212}
b_{ij}^q = \frac{\alpha_{i-1}^q + (i-1) \beta_{i-1}^q}{(\alpha_i^q - \alpha_j^q) + (i \beta_i^q -
	j \beta_j^q) + (\gamma_i^{*q} - \gamma_j^{*q})} b_{i-1,j}^q,
\ j=0,1,\dots,i-1,
\end{equation}
and the relationships for the coefficients $c_{ij}$ are ($i=1,\dots$; ):
\begin{equation}
c_{ij} = \frac{\alpha_{i-1}^B + (i-1) \beta_{i-1}^B}{(\alpha_i^B - \alpha_j^A) + (i \beta_i^B -
	j \beta_j^A) + (\gamma_i^{*B} - \gamma_j^{*A})} c_{i-1,j},\ j=1,2,,\dots,i-1,\nonumber 
\end{equation}
\begin{equation}\label{213}
c_{0j}=\frac{\delta_n}{\alpha_0^B+\gamma_0^{*B}-(\alpha_j^A+j\beta_j^A+\gamma_j^{*A})}b_{nj}^{A}, \ j=1,2,\dots,n 
\end{equation}
Further on in this study, we shall omit the upper index $q$. We shall use it where it is necessary to specify the exact channel.
\par
From Eq.(\ref{211}) one obtains
\begin{eqnarray}\label{214}
x_i^{*} = \frac{\prod \limits_{j=0}^{i-1}[\alpha_{i-j-1}+(i-j-1)\beta_{i-j-1}]}{
	\prod \limits_{j=0}^{i-1} \alpha_{i-j} + (i-j) \beta_{i-j} + \gamma_{i-j}^{*}} x_0^{*}; \ \ \
x^* = \sum \limits_{i=0}^\infty x_i^* = \nonumber \\
x_0^* \left \{ 1+ \sum \limits_{i=1}^\infty \frac{\prod \limits_{j=0}^{i-1}[\alpha_{i-j-1}+(i-j-1)\beta_{i-j-1}]}{
	\prod \limits_{j=0}^{i-1} \alpha_{i-j} + (i-j) \beta_{i-j} + \gamma_{i-j}^{*}} \right \}
\end{eqnarray}
The form of the corresponding stationary distribution $y_i^{*} = x_i^{*}/x^{*}$  (where $x^{*}$ is the amount of the substance in all of the cells of the arm of the channel) is
\begin{equation}\label{215}
y_0^* = \frac{1}{1+ \sum \limits_{i=1}^\infty \frac{\prod \limits_{j=0}^{i-1}[\alpha_{i-j-1}+(i-j-1)\beta_{i-j-1}]}{
		\prod \limits_{j=0}^{i-1} \alpha_{i-j} + (i-j) \beta_{i-j} + \gamma_{i-j}^{*}}}; \ \ \
y_i^{*} = \frac{\frac{\prod \limits_{j=0}^{i-1}[\alpha_{i-j-1}+(i-j-1)\beta_{i-j-1}]}{
		\prod \limits_{j=0}^{i-1} \alpha_{i-j} + (i-j) \beta_{i-j} + \gamma_{i-j}^{*}}}{1+ \sum \limits_{i=1}^\infty \frac{\prod \limits_{j=0}^{i-1}[\alpha_{i-j-1}+(i-j-1)\beta_{i-j-1}]}{
		\prod \limits_{j=0}^{i-1} \alpha_{i-j} + (i-j) \beta_{i-j} + \gamma_{i-j}^{*}}}, i=1,2,3,\dots
\end{equation}
To the best of our knowledge the distribution presented by Eq.(\ref{215}) was not discussed up to now outside our research group. Fig 2.  visualizes several cases of the distribution described by Eq.(\ref{215}).  Let us show that this distribution contains as particular cases several famous distributions, e.g., Waring distribution, Zipf distribution, and Yule-Simon distribution. In order to do this we consider the particular case when $\beta_i\ne 0$ and write $x_i$ from Eq.(\ref{214}) as follows
\begin{equation}\label{216}
x_i^{*} = \frac{\prod \limits_{j=0}^{i-1} b_{i-j} [k_{i-j-1} + (i-j-1)]}{\prod \limits_{j=0}^{i-1} [k_{i-j} + a_{i-j} + (i-j)]} x_0^{*}
\end{equation}
where $k_i = \alpha_i/\beta_i$; \ \ $a_i = \gamma_i^{*}/\beta_i$; \ \ $b_i = \beta_{i-1}/\beta_i$. The form of the corresponding stationary distribution $y_i^{*} = x_i^{*}/x^{*}$ is
\begin{equation}\label{217}
y_i^{*} = \frac{\prod \limits_{j=0}^{i-1} b_{i-j} [k_{i-j-1} + (i-j-1)]}{\prod \limits_{j=0}^{i-1} [k_{i-j} + a_{i-j} + (i-j)]} y_0^{*}
\end{equation}
Let us now consider the particular case where $\alpha_i = \alpha$ and $\beta_i = \beta$ for $i=0,1,2,\dots$. Then from Eqs.(\ref{216}) and (\ref{217}) one obtains
\begin{equation}\label{218}
x_i^* = \frac{[k+(i-1)]!}{(k-1)! \prod \limits_{j=1}^i (k+j+a_j)} x_0^*
\end{equation}
where $k = \alpha/\beta$ and $a_j=\gamma^*_j/\beta$. The form of the corresponding stationary distribution $y_i^* = x_i^*/x^*$ is
\begin{equation}\label{219}
y_i^* = \frac{[k+(i-1)]!}{(k-1)! \prod \limits_{j=1}^i (k+j+a_j)} y_0^*
\end{equation}
\begin{figure}[!htb]
	\centering
	\includegraphics[scale=.75]{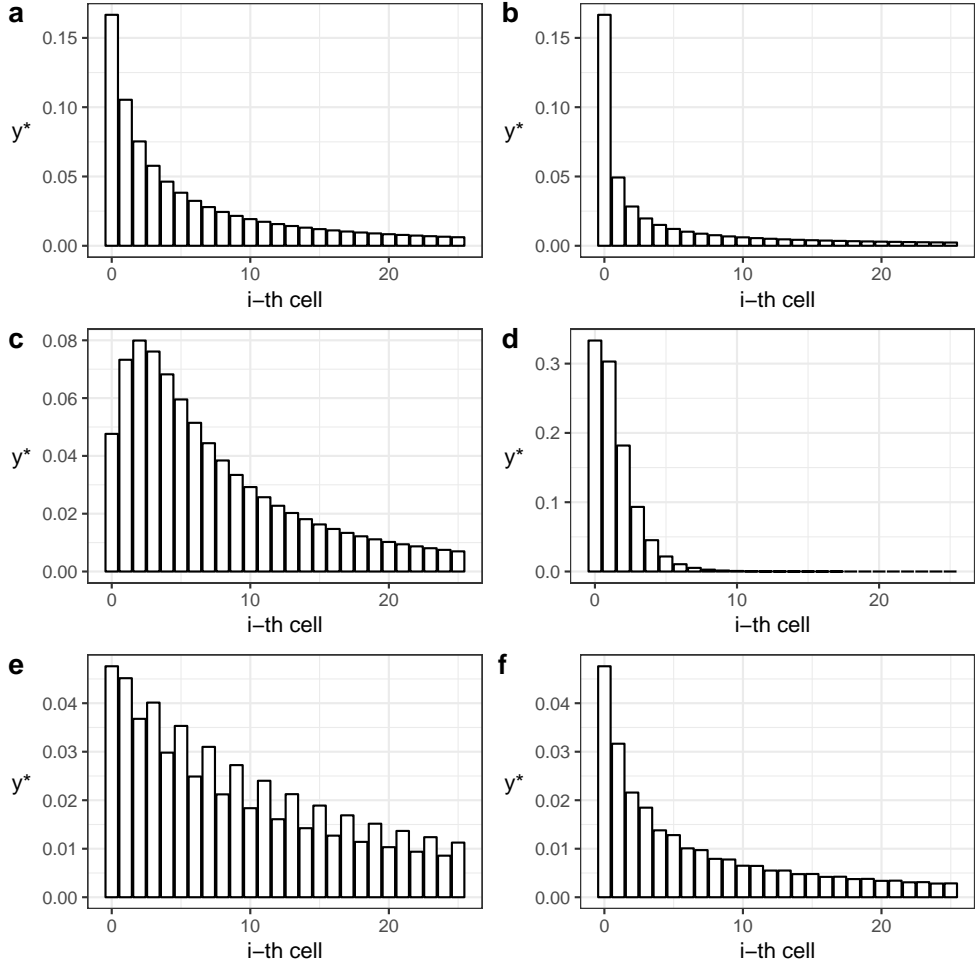}
	\caption{The distribution described by the Eqs.(\ref{215})  for the case of $\delta_i=0$ and for different values of the three parameters - $\alpha_i, \beta_i$ and $\gamma_i$. The figures show the influence of different choices of the parameters on the shape of the distribution. Each figure on the right column (2a),(2c) and (2d) shows the distribution for fixed $\beta_i=0.01,\ \gamma_i=0.01$ and various variable values of the parameter $\alpha_i$.  The figures on the right column show the distribution from the left with changed one of the parameters $\beta_i$ or $\gamma_i$. Figures~(2a) and (2b): $\alpha_i=0.05+0.1i/(i+10), \  \gamma_i=0.01$, $\beta_i=0.01$ for (2a) and  $\beta_i=0.1$ for (2b). Figures (2c) and (2d): $\alpha_i=0.2-0.18i/(i+1),\ \beta_i=0.01,\ \gamma_i=0.01$ for (2c) and $\gamma_i=0.1$ for (2d). Figures (2e) and (2f): $\alpha_i=0.2+0.1(-1)^i i/(i+10), \ \gamma_i=0.01,\ \beta_i=0.01$ for (2e) and $\beta_i=0.1$ for (2f). The shape of distribution is sensitive to differences of the values of parameters $\alpha_i, \beta_i$ and $\gamma_i$.}
	\label{fig:1}
\end{figure}
Let us consider the particular case where $a_0 = \dots = a_N$. In this case the distribution from Eq.(\ref{219}) is reduced to the distribution:
\begin{eqnarray}\label{220}
P(\zeta = i) &=& P(\zeta=0) \frac{(k-1)^{[i]}}{(a+k)^{[i]}}; \ \ k^{[i]} = \frac{(k+i)!}{k!}; \ i=1, 2, \dots 
\end{eqnarray}
$P(\zeta=0)=y_0^* = x_0^*/x^*$ is the percentage of substance that is located in the first cell of the channel. Let this percentage be 
\begin{equation}\label{221}
y_0^* = \frac{a}{a+k}
\end{equation}
The case described by Eq.(\ref{221}) corresponds to the situation where the amount of substance in the first cell is proportional of the amount of substance in the entire channel. In this case Eq.(\ref{219}) is reduced to:
\begin{eqnarray}\label{222}
P(\zeta = i) &=& \frac{a}{a+k} \frac{(k-1)^{[i]}}{(a+k)^{[i]}}; \ \ k^{[i]} = \frac{(k+i)!}{k!}; \ i=1, 2, \dots 
\end{eqnarray}
Let us denote $\rho = a$ and $k=l$. The distribution (\ref{222}) is exactly the Waring distribution (probability distribution of non-negative integers named after Edward Waring - the 6th Lucasian professor of Mathematics in Cambridge from the 18th century)  \cite{varyu1} - \cite{varyu3}
\begin{equation}\label{223}
p_l = \rho \frac{\alpha_{(l)}}{(\rho + \alpha)_{(l+1)}}; \
\alpha_{(l)} = \alpha (\alpha+1) \dots (\alpha+l-1)
\end{equation}
Waring distribution may be written also as follows
\begin{eqnarray}\label{224}
p_0 = \rho \frac{\alpha_{(0)}}{(\rho + \alpha)_{(1)}} = \frac{\rho}{\alpha + \rho}; \ \ 
p_l = \frac{\alpha+(l-1)}{\alpha+ \rho + l}p_{l-1}.
\end{eqnarray}
The mean $\mu$ (the expected value)  and the variance $V$ of the Waring distribution are
\begin{equation}\label{225}
\mu = \frac{\alpha}{\rho -1} \ \textrm{if} \ \rho >1; \ \
V = \frac{\alpha \rho (\alpha + \rho -1)}{(\rho-1)^2(\rho - 2)} \
\textrm{if} \ \rho >2
\end{equation}
$\rho$ is called the tail parameter as it controls the tail of the Waring distribution. Waring distribution contains various distributions as particular cases. Let $i \to \infty$ Then the Waring distribution is reduced to  the frequency form of the Zipf distribution \cite{chen}
\begin{equation}\label{226}
p_l \approx \frac{1}{l^{(1+\rho)}}.
\end{equation}
If $\alpha \to 0$ the Waring distribution is reduced to the Yule-Simon distribution \cite{simon} 
\begin{equation}\label{227}
p(\zeta = l \mid \zeta > 0) = \rho B(\rho+1,l)
\end{equation}
where $B$ is the beta-function. 
\section { Stationary regime of functioning of the third arm of the channel} \label{sec:num111}
Let us now consider the stationary regime of functioning of the third arm of the channel ($C$). Here we shall denote as $0$-th node the node where the third arm of the channel splits from the second arm ($B$). We assume that an amount $x^C$ of the substance  becomes distributed among the cells of the third arm of the channel and this substance can move fromone cell to another cell. Let $x_i^C$ be the amount of the substance in the $i$-th cell. Then
\begin{equation}\label{228}
x^C = \sum \limits_{i=0}^\infty x_i^C
\end{equation}
We shall use the upper index indicating the arm only where it is needed. The fractions $y_i = x_i/x$ can be considered as probability values of distribution of a discrete random variable $\zeta$: 
$y_i = p(\zeta = i), \ i=0,1, \dots$.
The content $x_i$ of any cell may change due to the same three processes that govern the motion of the substance in the first and second arms of the channel. The process of the motion of the substance is continuous in the time. Then the process can be modeled mathematically by the system of ordinary differential equations:
\begin{eqnarray} \label{230}
\frac{dx_0}{dt} &=& s-f_0-g_0; \nonumber \\
\frac{dx_i}{dt} &=& f_{i-1} - f_i - g_i, \ i=1,2,\dots.
\end{eqnarray}
The  relationships for the quantities of the above equations are ($\alpha, \beta, \gamma_i$ are constants)
\begin{eqnarray}\label{231}
s &=& \delta_m^B x_m^B;   \nonumber \\
f_i &=& (\alpha_i + \beta_i i) x_i; \ \ \ \alpha_i >0, \ \beta_i \ge 0  \nonumber \\
g_i &=& \gamma_i x_i; \ \ \ \gamma_i \ge 0 
\end{eqnarray}
Let's note that $\delta_n^A \ne \delta_m^B$ in general. Thus the system of equations for the motion of the substance in this arm of the channel is
\begin{eqnarray} \label{232}
\frac{dx_{0}}{dt}&=&\delta_m^B x_m^B-\alpha_0 x_0- \gamma_0 x_0  \nonumber \\
\frac{dx_{i}}{dt}&=&[\alpha_{i-1}+(i-1)\beta_{i-1}]x_{i-1}-
(\alpha_{i}+ i \beta_{i}+\gamma_{i})x_{i};\ \ i=1,2,\dots 
\end{eqnarray}
In this we shall discuss the situation in which the stationary state is established in the entire channel (in the three arm of the channel). In this case $x_m^B \to x_m^{*B}$; $\frac{dx_{0}}{dt} \to 0$ and $\frac{dx_{i}}{dt} \to 0$. Then
\begin{eqnarray}\label{233}
x_0^* = \frac{\delta_m^B x_m^{*B}}{\alpha_0 + \gamma_0}; \ \
x_i^* = \frac{\alpha_{i-1} + (i-1) \beta_{i-1}}{\alpha_i + i \beta_i + \gamma_i} x_{i-1}^*, \ i=1,2,\dots
\end{eqnarray} 
From the second of Eqs.(\ref{233}) one obtains
\begin{equation}\label{234}
x_i^* = \frac{\prod \limits_{j=0}^{i-1}[\alpha_{i-j-1}+(i-j-1)\beta_{i-j-1}]}{
	\prod \limits_{j=0}^{i-1} [\alpha_{i-j} + (i-j) \beta_{i-j} + \gamma_{i-j}]} x_0^*
\end{equation}
Considering Eq. (\ref{29}) and Eq. (\ref{214}),  Eq. (\ref{234}) takes the form
\begin{eqnarray}\label{235}
x_i^{*C} &=&\frac{\prod \limits_{j=0}^{i-1}[\alpha_{i-j-1}^C+(i-j-1)\beta_{i-j-1}^C]}{\prod \limits_{j=0}^{i-1} [\alpha_{i-j}^C + (i-j) \beta_{i-j}^C + \gamma_{i-j}^C]} \times \frac{\prod \limits_{j=0}^{n-1}[\alpha_{n-j-1}^B+(n-j-1)\beta_{n-j-1}^B]}{\prod \limits_{j=0}^{n-1}[ \alpha_{n-j}^B + (n-j) \beta_{n-j}^B + \gamma_{n-j}^{*B}]} \times \nonumber \\ 
&&\frac{\delta_n^A \delta_m^{B}}{(\alpha_0^B + \gamma_0^{*B})(\alpha_0^C + \gamma_0^C)}  x_n^{*A}
\end{eqnarray}
The form of the corresponding stationary distribution $y_i^{*C} = x_i^{*C}/x^{*C}$ (where $x^*$ is the amount of the substance in all of the cells of third arm of the channel) is
\begin{equation}\label{236}
y_i^{*C} = \frac{\prod \limits_{j=0}^{i-1}[\alpha_{i-j-1}+(i-j-1)\beta_{i-j-1}]}{
	\prod \limits_{j=0}^{i-1} [\alpha_{i-j} + (i-j) \beta_{i-j} + \gamma_{i-j}]} y_0^{*C}
\end{equation}
\par
Let us see the condition for arising of the Waring distribution in the third arm of the channel. In order to do this we consider the particular case when $\beta_i \ne 0$ and write $x_i$ from Eq.(\ref{234}) as follows
\begin{equation}\label{237}
x_i^{*C} = \frac{\prod \limits_{j=0}^{i-1} b_{i-j} [k_{i-j-1} + (i-j-1)]}{\prod \limits_{j=0}^{i-1} [k_{i-j} + a_{i-j} + (i-j)]} x_0^{*C}
\end{equation}
where $k_i = \alpha_i/\beta_i$; $a_i = \gamma_i/\beta_i$;  $b_i = \beta_{i-1}/\beta_i$. The form of the corresponding stationary distribution $y_i^{*C} = x_i^{*C}/x^{*C}$ is
\begin{equation}\label{238}
y_i^{*C} = \frac{\prod \limits_{j=0}^{i-1} b_{i-j} [k_{i-j-1} + (i-j-1)]}{\prod \limits_{j=0}^{i-1} [k_{i-j} + a_{i-j} + (i-j)]} y_0^{*C}
\end{equation}
Let us now consider the particular case where $\alpha_i = \alpha$ and $\beta_i = \beta$ for $i=0,1,2,\dots$. Setting in addition $k = \alpha/\beta$ and $a_j=\gamma_j/\beta$. We obtain the stationary distribution $y_i^{*C} = x_i^{*C}/x^{*C}$ as
\begin{equation}\label{239}
y_i^{*C} = \frac{[k+(i-1)]!}{(k-1)! \prod \limits_{j=1}^i (k+j+a_j)} y_0^{*C}
\end{equation}
Let us consider the particular case where $a_0 = \dots = a_N$. In this case the distribution from Eq.(\ref{239}) is reduced to the distribution:
\begin{eqnarray}\label{240}
P(\zeta = i) &=& P(\zeta=0) \frac{(k-1)^{[i]}}{(a+k)^{[i]}}; \ \ k^{[i]} = \frac{(k+i)!}{k!}; \ i=1, 2, \dots 
\end{eqnarray}
$P(\zeta=0)=y_0^{*C} = x_0^{*C}/x^{*C}$ is the percentage of substance that is located in the first cell of the channel. Let this percentage be 
\begin{equation}\label{241}
y_0^{*C} = \frac{a}{a+k}
\end{equation}
In this case Eq.(\ref{240}) is reduced to the  distribution:
\begin{eqnarray}\label{242}
P(\zeta = i) &=& \frac{a}{a+k} \frac{(k-1)^{[i]}}{(a+k)^{[i]}}; \ \ k^{[i]} = \frac{(k+i)!}{k!}; \ i=1, 2, \dots 
\end{eqnarray}
We arrive again to the Waring distribution.
\section{Relations between distribution of migrants and the economics}
Migrants earn money and send part of them to the home countries. let us discuss a migration channell that has reached stationary state of motion of migrants and  let the
mean amount of money  send by a migrant per unit time  is $M_i$. Then the amount of money sent by the migrants from the $i$-th country is $S_i = M_i x_i^*$. The amount of money sent home by the migrants from country $r-1$ to country $r_2$ is
\begin{equation}\label{ea1}
S_{r_1,r_2} = \sum \limits_{i=r_1}^{r_2} M_i x_i^* =
\sum \limits_{i=r_1}^{r_2} M_i \frac{\prod \limits_{j=0}^{i-1}[\alpha_{i-j-1}+(i-j-1)\beta_{i-j-1}]}{
	\prod \limits_{j=0}^{i-1} \alpha_{i-j} + (i-j) \beta_{i-j} + \gamma_{i-j}^{*}} x_0^{*}
\end{equation}
The relation between the amount of money sent from the different segments of the channel is
\begin{equation}\label{ea2}
R = \frac{S_{r_1,r_2}}{S_{r_3,r_4}} = \frac{\sum \limits_{i=r_1}^{r_2} M_i \frac{\prod \limits_{j=0}^{i-1}[\alpha_{i-j-1}+(i-j-1)\beta_{i-j-1}]}{
		\prod \limits_{j=0}^{i-1} \alpha_{i-j} + (i-j) \beta_{i-j} + \gamma_{i-j}^{*}}}{\sum \limits_{i=r_3}^{r_4} M_i \frac{\prod \limits_{j=0}^{i-1}[\alpha_{i-j-1}+(i-j-1)\beta_{i-j-1}]}{
		\prod \limits_{j=0}^{i-1} \alpha_{i-j} + (i-j) \beta_{i-j} + \gamma_{i-j}^{*}}}
\end{equation}
Of special interest are the relations $R$ where $r_4 = \infty$ and $r_3=r_2+1$. In this case one obtains the ratio of the money sent from some segment of the channel and the money send from the the rest of the channel after this segment. Other relations can be obtained too and of special interest is the theory connected to finite-size channels which results will be reported elsewhere.



\begin{thebibliography}{99}
\bibitem{a1}
M. Ausloos, A. Gadomski, N. K. Vitanov. Physica Scripta \textbf{89}, 108002 (2014).
\bibitem{dx3}
Z. I. Dimitrova. Journal of Theoretical and Applied Mechanics \textbf{45}, 79 -- 92 (2015).
\bibitem{d2}
Z. I. Dimitrova, N. K.  Vitanov. Journal of Physics A: Mathematical and General \textbf{34}, 7459 -- 7473 (2001).
\bibitem{d3}
Z. I. Dimitrova, N. K. Vitanov. Physica A \textbf{300}, 91 -- 115 (2001).
\bibitem{d4}
Z. I. Dimitrova,  N. K.Vitanov. Theoretical Population Biology \textbf{66}, 1 -- 12 (2004).
\bibitem{k1}
H. Kantz, D. Holstein, M. Ragwitz, N. K. Vitanov. Physica A \textbf{342}, 315 -- 321 (2004). 
\bibitem{kal1}
K. Vitanov, M. Slavtchova-Bojkova. Annual of Sofia University "St. Kliment Ohridski", Faculty of Mathematics and Informatics, \textbf{104}, 193 -- 200 (2018)
\bibitem{vx1}
N. K. Vitanov, Z. I. Dimitrova.  Communications in Nonlinear Science and Numerical Simulation \textbf{15}, 2836 -- 2845 (2010).
\bibitem{vx1a}
N. K. Vitanov, Z. I. Dimitrova, H.  Kantz. 
Applied Mathematics and Computation \textbf{216}, 2587 -- 2595 (2010).
\bibitem{vx1d}
N. K. Vitanov, Z. I. Dimitrova, K. N. Vitanov. Applied Mathematics and Computation \textbf{269}, 363 -- 378 (2015)
\bibitem{vx1e}
N. K. Vitanov, Z. I. Dimitrova, H. Kantz
Applied Mathematics and Computation \textbf{219}, 7480 -- 7492 (2013)
\bibitem{vx3}
N. K. Vitanov, N. P. Hoffmann, B. Wernitz.  Chaos, Solitons \& Fractals, \textbf{69}, 90 -- 99 (2014).
\bibitem{elena1}
E. V. Nikolova, V. K. Kotev, G. S. Nikolova. EMBEC \& NBC 2017. EMBEC 2017, NBC 2017. IFMBE Proceedings, \textbf{65}, Springer, Singapore, 209--212 (2018).
\bibitem{elena2}
E. V. Nikolova. AIP Conference Proceedings \textbf{1978}, No.1, 470050 (2018).
\bibitem{elena3}
T. B. Ivanov, E. V. Nikolova.  Advanced Computing in Industrial Mathematics. Studies in Computational Intelligence \textbf{681}, Springer, Cham, 61--74 (2017).
\bibitem{elena4}
E. Nikolova, T. Ivanov. Series on Biomechanics \textbf{29},  78--84 (2015).
\bibitem{elena5}
S. Tabakova, E. Nikolova, S. Radev. AIP Conference Proceedings \textbf{1629}, No.1, 336--343 (2014).
\bibitem{elena6}
E. Nikolova. Compt. rend. Acad. bulg.  Sci., \textbf{65}, 33--40 (2012).
\bibitem{elena7}
V. Petrov, E. Nikolova, O. Wolkenhauer. IET systems biology \textbf{1}, 
No. 1, 2 -- 9 (2007)
\bibitem{elena8}
V. Petrov, E. Nikolova, J. Timmer. Journal of Theoretical and Applied Mechanics \textbf{34}, 55 -- 78 (2004).
\bibitem{elena9}
E. Nikolova, E. Goranova, Z. Dimitrova. Compt. rend. Acad. bulg.  Sci. \textbf{69}, 1213--1222 (2016).
\bibitem{elena10}
I. Jordanov, E. Nikolova. Journal of Theoretical and Applied Mechanics \textbf{43}, 69--76 (2013).
\bibitem{elena11}
E. Nikolova, V. Petrov. Compt. rend. Acad. bulg.  Sci. \textbf{63}, 1421--1428 (2010).
\bibitem{elena12}
E. Nikolova, V. Petrov, I. Edissonov. Journal of Theoretical and Applied Mechanics \textbf{41}, 83--92 (2011).
\bibitem{elena13}
I. Edissonov, E. Nikolova, S. Ranchev. Compt. rend. Acad. bulg.  Sci.  \textbf{61}, 1401--1406 (2008).
\bibitem{elena14}
E. Nikolova. Compt. rend. Acad. ́bulg.  Sci.  \textbf{59}, 143--150 (2006).
\bibitem{elena15}
E. V. Nikolova, J. Timmer, V. G. Petrov. Series on Biomechanics \textbf{24}, 79--100 (2009).
\bibitem{vx4}
N. K. Vitanov, Z. I. Dimitrova, K. N. Vitanov.  Applied Mathematics and Computation \textbf{269}, 363 -- 378 (2015).
\bibitem{zlat1}
Z.I. Dimitrova. M. Ausloos. Open Physics \textbf{13}, 218 -- 225 (2015).
\bibitem{zlat2}
Z. Dimitrova.  Journal of Theoretical and Applied Mechanics \textbf{42}, No. 3, 3 -- 22 (2012).
\bibitem{vx5}
N. K. Vitanov, M. Ausloos. Journal of Applied Statistics, \textbf{42}, 2686 -- 2693 (2015).
\bibitem{vx6}
N. K. Vitanov, Z. I.  Dimitrova, T. I. Ivanova. Applied Mathematics and Computation \textbf{315},  372 -- 380 (2017) .
\bibitem{albert}
R. Albert, A. -L. Barabasi. Rev. Mod. Phys. \textbf{74}, 47 -- 97 (2002). 
\bibitem{bok}
S. Boccaletti, V. Latora, Y. Moreno, M. Chavez, D. U. Hwang. Physics Reports, \textbf{424},  175 -- 308 (2006).
\bibitem{vk1}
N. K. Vitanov. {\sl Science dynamics and research production: Indicators, indexes, statistical laws and mathematical models} (Springer, Cham, 2016).
\bibitem{jord1}
I. P. Jordanov. Compt. rend. Acad. bulg. Sci. \textbf{62} 33 -- 40 (2009).
\bibitem{vit3}
N. K. Vitanov, I. P. Jordanov, Z. I. Dimitrova. 
Communications in Nonlinear Science and Numerical Simulation \textbf{14}, 2379 -- 2388 (2009).
\bibitem{jord2}
I. P. Jordanov. Compt. rend. Acad. bulg. Sci. \textbf{61} 307 -- 314 (2008).
\bibitem{vit4}
N. K. Vitanov, I. P. Jordanov, Z. I. Dimitrova. Applied Mathematics and Computation \textbf{215}, 2950 - 2964 (2009).
\bibitem{vit1}
N. K. Vitanov, Z. I. Dimitrova, M. Ausloos. Physica A \textbf{389}, 4970 - 4980 (2010).
\bibitem{vit2}
N. K. Vitanov, M. Ausloos, G. Rotundo. Advances in Complex Systems \textbf{15} Supplement 1, Article number 1250049 (2012).
\bibitem{va12}
N. K. Vitanov, M. Ausloos. Knowledge epidemics and population dynamics models for describing idea diffusion, in \emph{Models of science dynamics},  edited by A. Scharnhorst, K. B{\"o}rner, P. van den Besselaar.  (Berlin, Springer, 2012),pp. 69 -- 125 .
\bibitem{vit5}
N. K. Vitanov, Z. I. Dimitrova, K. N. Vitanov. Computers \& Mathematics with Applications \textbf{66}, 1666 -- 1684 (2013).
\bibitem{vit5a}
N. K. Vitanov, K. Sakai, I. P. Jordanov, S. Managi, K. Demura
Physica A: Statistical Mechanics and its Applications \textbf{382}, 330 -- 335 (2007).
\bibitem{vit5b}
K. Sakai, S. Managi, N. K. Vitanov, K. Demura
Nonlinear dynamics, psychology, and life sciences \textbf{11}, 253 -- 265 (2007).
\bibitem{vit6}
N. K. Vitanov, K. N. Vitanov. Computers \& Mathematics with Applications \textbf{68}, 962 -- 971 (2013).
\bibitem{ff}
L. D. Ford, Jr., D. R. Fulkerson. \emph{Flows in networks} ()Princeton University Press, Princeton, NJ, 1962).
\bibitem{ahuja}
R.K. Ahuja, T. L. Magnanti, J. B. Orlin. \emph{Network flows. Theory, algorithms, and applications} (Prentice Hall, NJ, 1993).
\bibitem{todinov}
M. T. Todinov. \emph{Flow networks. Analysis and optimization of repairable flow networks, networks with disturbed flows, static flow networks and reliability networks} (Elsevier, Amsterdam, 2013).
\bibitem{ch1}
W.-K. Chan. \emph{Theory of nets: Flows in networks} (Wiley, New York, 1990).
\bibitem{everet}
E. S. Lee. Demography \textbf{3}, 47 -- 57 (1966).
\bibitem{armi}
R. Armitage.  Population Trends \textbf{43},  31-40 (1986).
\bibitem{ht}
J. R. Harris, M. P. Todaro.  The American Economic Review \textbf{60}, 126 -- 142 (1970).
\bibitem{smn}
J. H. Simon. \emph{The economic consequences of migration} (The University of Michigan Press, Ann Arbor, MI, 1999).
\bibitem{borj}
G. J. Borjas. International Migration Review \textbf{23}, 457 - 485 (1989).
\bibitem{will99}
F. J. Willekens. SA Journal of Demography \textbf{7}, 31 -- 43 (1999).
\bibitem{sg1}
A. Schubert, W. Gl{\"a}nzel. Scientometrics \textbf{6}, 149 -- 167 (1984).
\bibitem{vk4}
N. K. Vitanov, K. N. Vitanov, T. Ivanova. Box model of migration in channels of migration networks, in \emph{Advanced Computing in Industrial Mathematics}, edited by  K. Georgiev et al., Studies in Computational Intelligence No. 728, Springer, Berlin, (2018), pp. 203 - 215.
\bibitem{vk2}
N. K. Vitanov, K. N. Vitanov.  Physica A \textbf{490}, 1277 -- 1294 (2018).
\bibitem{vk3}
N. K. Vitanov, K. N. Vitanov. Physica A \textbf{509}, 635 -- 650    (2018).
\bibitem{rmr07}
J. Raymer. 
Environment and Planning A \textbf{39}, 985 -- 995 (2007).
	\bibitem{v1}
	N. K. Vitanov, K. N. Vitanov. Mathematical Social Sciences \textbf{80}, 108 -- 114 (2016).
	\bibitem{sg1}
	A. Schubert, W. Gl{\"a}nzel. Scientometrics \textbf{6} , 149 -- 167 (1984).
	\bibitem{varyu1}
	J. O. Irwin. Journal of the Royal Statistical Society \textbf{126}, 1 -- 44 (1963).
	\bibitem{varyu2}
	J. O. Irwin. Journal of the Royal Statistical Society \textbf{131}, 205 - 225 (1968). 
	\bibitem{varyu3}
	V. Diodato. \emph{Dictionary of Bibliometrics} (Haworth Press, Binghampton, NY, 1994).
	\bibitem{chen}
	W.-C. Chen. Journal of Applied Probability \textbf{17}, 611 -- 622 (1980).
	\bibitem{simon}
	H. A. Simon. Biometrica \textbf{42}, 425 -- 440 (1955). 
\end{thebibliography}
\end{document}